\documentstyle[aps,prl,twocolumn]{revtex} 
\begin{document}
\draft

\title{The moment method in general Nonlinear Schr\"odinger Equations}

\author{Juan J. Garc\'{\i}a-Ripoll and V\'{\i}ctor M.
  P\'erez-Garc\'{\i}a}

\address{Departamento de Matem\'aticas, Escuela T\'ecnica Superior de
  Ingenieros Industriales, \\
  Universidad de Castilla--La Mancha, 13071 Ciudad Real, Spain.}

\date{\today}

\maketitle
\begin{abstract}
  In this paper we develop a new approximation method valid for a wide
  family of nonlinear wave equations of Nonlinear Schr\"odinger type.
  The result is a reduced set of ordinary differential equations for a
  finite set of parameters measuring global properties of the
  solutions, named momenta. We prove that these equations provide
  exact results in some relevant cases and show how to impose
  reasonable approximations that can be regarded as a perturbative
  approach and as an extension of the time dependent variational method.
\end{abstract}

\pacs{PACS:
03.40.Kf,    
42.65.-k,    
03.75.-b    
}

\narrowtext


The solution of nonlinear wave equations representing physically
relevant phenomena is a task of the highest interest. However, it is
not possible to find exact solutions except for a few simple cases
where one is lucky to integrate the equations involved. In particular,
in the 70's some mathematical techniques were discovered which allowed
the integration of several relevant nonlinear wave equations
\cite{Zaharov}. So the development of rigorous approximation methods
is of interest in those cases where the equations are known to be
non-integrable.

One family of nonlinear wave equations with lots of practical
applications is that of Nonlinear Schr\"odinger Equations (NSE)
\cite{Vazquez,Kivshar}, which arise in plasma physics, biomolecule
dynamics, fundamentals of quantum mechanics, beam physics, etc., but
specially in the fields of Nonlinear Optics \cite{Hasegawa} and
Bose--Einstein condensation \cite{Dalfovo}. In the last two fields a
great variety of these equations appear involving different spatial
dimensionalities, nonlinear terms (saturation, polynomial, nonlocal,
losses, etc.) and number of coupled equations.

One common approximate theoretical approach to the analysis of the
dynamics involved in those problems is to assume a fixed shape for the
solution with a finite set of time dependent parameters so that the
dimensionality is made finite at the cost of loss of information on
the solution (many degrees of freedom are lost). This method receives
many denominations depending on the context: collective coordinate
technique, time-dependent variational method, equivalent particle
approach, energy balance equations, etc.

Although not explicitly stated most of those methods can be reduced to
a more elegant formulation which is the time dependent variational
technique, originally developed by Anderson \cite{Anderson83} for one
dimensional problems based on Ritz's optimization procedure.  This
approximate technique is a good tool to study the propagation of
distributions having simple shape. If the shape of the actual solution
is close to the trial function, the outcome of variational method will
be in good agreement with the real solutions, otherwise it may be very
rough or even fail \cite{Ankiewicz93}.  Despite of this fact the
technique has been used in many physical situations. In Nonlinear
Optics it has been applied to many problems some of them being listed
in Ref. \cite{Caglioti90}. The method has been applied to many other
physical problems where nonlinear wave equations (in particular NSEs)
arise including random perturbations \cite{random}, nonlocal equations
\cite{nonloc}, collapse phenomena \cite{Rasmussen94,PRA},
propagation and scattering of nonlinear waves \cite{Fei}, etc.  A
review of the application of the technique with emphasis on problems
with different scales (focused in condensed matter) is given in Ref.
\cite{Angel}.  This technique has been also used in the last years in
the framework of Bose-Einstein condensation (BEC) applications to
explain the low energy excitation spectrum of single \cite{PRL} and
double \cite{PRA} condensates, collapse dynamics \cite{PRA},
and many other problems \cite{finite,Juanjo}, etc.

In this letter we develop a completely different technique called the
moment method \cite{nombres}. This method is based on the definition
of several integral parameters whose evolution can be computed in
closed form and has been used to obtain exact results in particular
applications \cite{Porras,Perez95,resonan}. We provide here a general
framework for its application as well as several ways to treat it
systematically as a perturbative technique.

{\em The general NSE and moment equations.-} Let us consider the
$n$-dimensional NSE
\begin{equation}
\label{NLSE}
i\frac{\partial \psi }{\partial t}=-\frac{1}{2}\triangle \psi +V(\vec{r})\psi +g(|\psi |^{2},t)\psi -i\sigma (|\psi |^{2},t)\psi 
\end{equation}
where $\triangle = \sum_{k} \partial^2/\partial x_j^2$, and
$V(\vec{r})=\sum_k\frac{1}{2}\omega_k(t)^{2}(t)x_k^2$ is a time and
spatially dependent parabolic potential which has been included
because it is always present in BEC problems. To aid readability, we
will separate the solution in modulus and phase,
$\psi=\sqrt{\rho}e^{i\phi}$, and define an interaction energy density
$G(\rho)$ as $g=\partial G/\partial \rho$.  The nonlinear terms
will be analytic functions of $\rho$ such that
$g(\rho),G(\rho)\rightarrow 0,\, \rho \rightarrow 0.$

In Table \ref{moment-definitions} we define the so called momenta of 
$\psi$.  Some of them are related to the momenta of the distribution $\rho=|\psi|^2$, and they all
have a physical meaning (see Ref. \cite{Porras} for their
interpretation in Optics). The evolution
equations for these quantities are
\begin{mathletters}
\begin{eqnarray}
\label{central-equations}
\frac{dN}{dt} & = & -2\int \sigma \rho ,\\
\frac{dX_i}{dt} & = & V_i-2\int \sigma x_i\rho ,\\
\frac{dV_i}{dt} & = & -\omega _iX_i-2\int \sigma \frac{\partial \phi }{\partial x_i}\rho ,\\
\frac{dW_i}{dt} & = & B_i-2\int \sigma x_i^2\rho ,\\
\frac{dB_i}{dt} & = & 4K_i-2\omega _i^2W_i-2\int DG-4\int \sigma x_i\frac{\partial \phi }{\partial x_i}\rho ,\\
\frac{dK_i}{dt} & = & -\frac{1}{2}\omega _i^2B_i-\int DG\frac{\partial ^2\phi }{\partial x_i^2}\\
 & + & \int \sigma \left[ \sqrt{\rho }\frac{\partial ^2\sqrt{\rho }}{\partial x_i^2}-\rho \left( \frac{\partial \phi }{\partial x_{\hat{i}}}\right) ^2\right] ,\\
\frac{dJ}{dt} & = & \sum _i\int DG\frac{\partial ^2\phi }{\partial x_i^2}-2\int \sigma g\rho +\int \frac{\partial G}{\partial t}.
\end{eqnarray}
\end{mathletters}
Here $DG$ is a shorthand notation for $G(\rho)-g(\rho)\rho$. Through
this paper we will concentrate in the most common case 
$\sigma (|\psi|^2,t) = \sigma (t)$ for which the equations are
\begin{mathletters}
\label{disip}
\begin{eqnarray}
\frac{dN}{dt} & = & -2\sigma N, \label{disipa} \\
\frac{dX_i}{dt} & = & V_i-2\sigma X_i,\label{disipb} \\
\frac{dV_i}{dt} & = & -\omega _iX_i-2\sigma V_i, \label{disipc} \\
\frac{dW_i}{dt} & = & B_i-2\sigma W_i,\\
\frac{dB_i}{dt} & = & 4K_i-2\omega _i^2W_i-2\int DG-2\sigma B_i,\\
\frac{dK_i}{dt} & = & -\frac{1}{2}\omega _i^2B_i-\int DG\frac{\partial ^2\phi }{\partial x_i^2}-2\sigma K_i,\\
\frac{dJ}{dt} & = & \sum _i\int DG\frac{\partial ^2\phi }{\partial x_i^2}-2J+\int \frac{\partial G}{\partial t}.
\end{eqnarray}\end{mathletters}
As stated before, some of these laws can be found in other treatments
\cite{Kivshar,Porras} which mostly concentrate on particular cases or
treat them as basis for perturbation methods choosing one particular shape for 
the solution. From Eqs.
(\ref{disipb}-\ref{disipc}) we find \emph{exact} closed equations for
zeroth and first order momenta,
\begin{equation}
\frac{d^2X_i}{dt^2} = -\omega _i(t)X_i-2 \sigma (t)\frac{dX_i}{dt}-2\sigma (t)X_i.
\end{equation}
In conservative systems with any type of potential the classical
result of Quantum Mechanics $d\left\langle x_i\right\rangle/dt =
-\left\langle \partial V/\partial x_i\right\rangle$ is obtained as
described in \cite{Juanjo}.

Once Eqs. (\ref{disipa}-\ref{disipc}) are integrated one is left with
the problem of solving the remaining $3n+1$ equations. Typically these
equations do not form a closed set but involve integral quantities
which are not included in the definitions of the momenta. To close
them, i.e. to equal the numbers of equations and of unknowns, one must
either restrict the problem or impose some kind of approximation.

{\em Exact closure of moment equations.-} We have found only two
relevant cases in which the closure is exact for the rest of the
equations. Both simplified problems correspond to conservative,
$\sigma=0$, spherically symmetric potentials, $\omega _i=\omega(t)$
where
\begin{mathletters}
\label{radial-moments}
\begin{eqnarray}
\frac{dR}{dt} & = & B_{r},\\
\frac{dB_{r}}{dt} & = & 4K-2\omega ^2(t)R-2D,\\
\frac{dK}{dt} & = & -\frac{1}{2}\omega ^2(t)B_{r}-\frac{dJ}{dt},
\end{eqnarray}
\end{mathletters}
being $D=\int [G-g\rho]$ and $R=\sqrt{W}$ the radial width. The first
integrable case corresponds to $n=2$ and $G=U\rho ^2$. For it Eqs.
(\ref{radial-moments}) simplify to
\begin{equation}
\label{radial-ode}
\frac{d^2R}{dt^2}=-\omega(t) R + \frac{M}{R^{3}}.
\end{equation}
Here $M$ a constant that depends only on the initial data and
interaction strength $U$.  This equation has been used to prove the
existence of extended resonances in Ref. \cite{resonan}.

Another ample family of systems for which moment equations are closed
and exact is those with a time-independent interaction strength,
$\partial G/\partial t=0$ (the usual case), and a divergenceless
velocity distribution (given by the phase gradient)
$\text{div}\left(\nabla\phi\right) = \triangle \phi = 0.$
This condition imposes no restriction on the density distribution
$\rho$ and is automatically satisfied by the well known vortex-line
solutions, which in $n$ spatial dimensions read
\begin{mathletters}
\label{divergenceless}
\begin{eqnarray}
\psi & = & \rho _{B}(x_1,...,x_n; t)e^{i\phi_B},\\
\phi_B & = & \vec{\alpha }\vec{X}+ \arctan \frac{x_k-X_k}{x_l-X_l}.
\end{eqnarray}\end{mathletters}
Here $\vec{X}$ are free, time dependent parameters, and $\rho_0$ is
arbitrary. With those simple conditions we get an infinite number of
constants of evolution named ``supermomenta'', $Q(F) \equiv \int
F(\rho)$, built up from differentiable functions of the modulus,
$F(\rho)$, that satisfy the regularity conditions. In these cases one
can prove that $D$ is a constant and that Eqs. (\ref{radial-moments})
become equivalent to Eq. (\ref{radial-ode}).

{\em Uniform divergence approximation.-} Intuition dictates that the
zero-Laplacian phase condition is related
to (a) the configuration of the cloud, i.e. $\rho$, does not change,
(b) the soliton or the wavepacket is either stationary or at most
suffers \emph{displacements and rotations} and (c) all of the
``supermomenta'' depend solely on the norm, $Q(F)=f(N)$. General
solutions have nonzero divergence of the velocity field, thus the
zeroth order approximation, $\triangle \phi=0$, fails to describe the
system. The next possibility is a first order approximation in which
the Laplacian of the phase is uniform. As we will see, this extension
now allows for \emph{changes in the shape} of the cloud and introduces
three new independent variables in the supermomenta $Q(F)$. Mathematically,
the  first order approximation the phase is
\begin{equation}
\label{parabolic-fit}
\phi = \phi_B(\vec{x},t) + \sum _j\beta_j(t)x_j^2,
\end{equation}
where $\phi_B(\vec{x},t)$ is any function satisfying $\triangle \phi_B =0$. 
This approximation 
will be called uniform divergence approximation on the phase in what
follows.  A limited version of this approximation that uses a linear
function (which has zero divergence) in place of $\phi_B$ was first applied
to radially symmetric problems in Ref. \cite{Perez95} and to the study of
resonances in general 3D forced NSE problems in Refs. \cite{Juanjo,resonan}. In that case, were
$\phi({\vec{x}},t) = \phi_0(t) + {\vec{\alpha}}{\vec{x}} + \sum _j\beta_j(t)x_j^2$
it is evident that the approximation consists on a Taylor expansion of the phase or even better a 
polynomial fitting with time dependent parameters. 
Here we provide a general framework for the application of the technique to
arbitrary NSE problems.  In a certain sense this is a generalization of the
usual time dependent variational method but now {\em no assumption on the
  shape of the amplitude of the wave is needed} and the phase has a free is
approximated by a least-squares type fitting with time-dependent parameters
and only very general restrictions on the phase.

Let us first take the case with $\phi_B=\vec{\alpha}\vec{x}$. We will assume
for simplicity that the nonlinearity is or can be approximated by a polynomial
\begin{equation}
\label{polynomial-nonlinearity}
G(\rho)=\sum _{k}\alpha _{k}\rho ^{k}.
\end{equation}
The dissipative terms can be removed from the equations by rescaling the
solution with $\gamma =\int _{0}^{t}\sigma (t')dt',$ so that $\tilde{\rho
  }=e^{\gamma }\rho$. We will denote the momenta obtained using
$\tilde{\rho}$ by a tilde, as in $\tilde{W}_i$. It is important to stress
that Eqs. (\ref{disip}) do not close immediately and it is necessary to
study the monomial supermomenta
\begin{equation}
\tilde{Q}^{(m)}=\int \tilde{\rho}^m d^nx.
\end{equation}
Their evolution laws are
\begin{equation}
\frac{d\tilde{Q}^{(m)}}{dt}=-(m-1)\sum_i\beta _i\tilde{Q}^{(m)}.
\end{equation}
Since the parameters in the phase can be expressed as
\begin{equation}
\beta _i=\frac{d}{dt}\log \sqrt{\tilde{W}_i},
\end{equation}
one obtains a closed form for each of the supermomenta
\begin{equation}
\label{supermoments}
\tilde{Q}^{(m)}=C^{(m)}\left( \sqrt{\tilde{W}_1\ldots \tilde{W}_d}\right) ^{-m+1},
\end{equation}
where the constants $C^{(m)}$ must be determined from initial data.  We
can use this property to estimate all the integrals in which $G(\rho)$
appears, as a function of powers of the mean square widths.  Using
this, and defining as before the natural widths,
$\tilde{R}_i=\sqrt{\tilde{W}_i}$, in the general nonsymmetric case we
arrive to
\begin{equation}
\label{asymmetric-ode}
\frac{d^2\tilde{R}_i}{dt}=-\omega _i\tilde{R}_i+\frac{M_i}{\tilde{R}_i^3}+\frac{{\cal G}(\tilde{R}_{1}\ldots \tilde{R}_{d},t)}{\tilde{R}_{1}}
\end{equation}
where $M_i$ is a constant to be determined from the initial data, and
${\cal G}(R_{1}\ldots R_{d},t)$ is a function that may be calculated
from Eqs. (\ref{polynomial-nonlinearity}) and (\ref{supermoments}).
Finally, to interpret these results one must remember that the actual
width of the cloud is actually $R_i(t)=\tilde{R}_i(t)e^{-2\gamma}$.

Eq. (\ref{asymmetric-ode}) means that given an initial data it is
possible to compute the evolution of the width (and all the momenta of
the initial datum) provided the parabolic phase is a good description
for the solution. The method is very powerful as it accounts for the
evolution of any solution that can be described in terms of a finite
number of momenta (since higher order momenta are functions of lower
order ones). Though much more general than the collective coordinate
method the one presented here is simpler to apply and extend since it
only involves computing the integrals in Eqs.  (\ref{disip}).

We must also remark that the phase is not restricted to be a polynomial in 
$\vec{r}$ as in Ref. \cite{Perez95}. 
Instead one still obtains information when the solution has many other forms.
 For example, if phase is of type (\ref{divergenceless}), then
one can combine the two widths from the plane on which the vorticity is
present, $W_k$ and $W_l$, into a radial one, $R\equiv W_k+W_l$ and 
 the evolution of $R$ can be studied together with the widths from the
remaining spatial directions.

{\em Independent moment approximation.-} We have seen that the uniform
divergence condition leads to a closed expression for every
$Q^{(m)}$ in terms of the widths $W_i$. Although this approach
is more powerful than the usual time dependent ansatz it is possible
to improve the moment method to higher precision. The
basic idea is to assume that only a finite set of momenta are
independent, the number of those independent momenta being related to
the accuracy of the solution and finding expressions for higher order
momenta as functions of lower order ones.  For instance if only the
$\{N,W_i\}$ momenta are truly independent (as it occurs with the
uniform divergence approximation) it is possible to find expressions for
the rest of the magnitudes in terms of these momenta. By scaling
the wave with respect to one of the coordinates
\begin{equation}
\psi (x_1,\ldots ,x_i,\ldots )\rightarrow \frac{1}{\sqrt{1+\epsilon }}\psi (x_1,\ldots ,\frac{x_i}{1+\epsilon },\ldots ),
\end{equation}
and relating the first order changes in $Q^{(m)}$ and $W_k$ one
arrives to
\begin{equation}
-(m-1)Q^{(m)}=\sum_k 2\frac{\partial Q^{(m)}}{\partial W_k}W_k.
\end{equation}
This equation has a solution of the form reflected in Eq.
(\ref{supermoments}) and also similar expressions may be derived for
the rest of the unknown integrals in Eq. (\ref{central-equations}).
Thus, in the first order case discussed here the method can provide a
closed set of equations for any type of nonlinear term. In principle
the procedure could be extended to higher precision approximation of
the evolution of initial data. Work on this point is in progress and
will be reported in detail elsewhere.

Finally we would like to point out that Eqs. (\ref{disip}) can be used
straightforwardly by replacing $\psi (\vec{r})$ with an appropriate
ansatz. In the conservative case, $\sigma =0$,
this procedure is equivalent to Ritz's optimization procedure, with
the advantages that one needs not build a huge Lagrangian integral,
and that our central equations (\ref{central-equations}) can also be
applied to dissipative systems that lack a Lagrangian density at all.
It must also be remarked that unless the ansatz has a phase different
from (\ref{parabolic-fit}), one will always arrive
to Eq. (\ref{asymmetric-ode}).

In conclusion we have presented the moment method in a general
framework and discussed under which conditions it leads to closed
equations for a finite set of parameters. The uniform divergence ansatz
has been introduced as a way to improve usual collective coordinate
methods and obtain a closed set of equations for any NSE
with polynomial nonlinearity and linear dissipation (with or without 
a parabolic time dependent spatial potential). The second method takes the
zeroth, first and second order momenta to be the only independent ones
and builds analytical approximations for other magnitudes. It is an
extension of the uniform divergence approximation that seems to work for
non polynomic self-interaction energies.

This work has been partially supported by CICYT under grant PB96-0534.

\begin{table}
\caption{Definitions of the momenta and physical interpretation.}
\begin{tabular}{ll}
Definition & Interpretation \\ \hline \\
$N = \int \rho$ & Norm, number of particles or intensity \\
$X_i = \int x_i\rho$ & Center of mass or of energy \\
$V_i = \int \frac{\partial \phi }{\partial x_i}\rho$ & Speed of center of mass\\
$W_i = \int x_i^2\rho$ & Widths \\
$B_i = 2\int x_i\frac{\partial \phi }{\partial x_i}\rho$ & Speed of growth\\
$K_i = -\frac{1}{2}\int \bar{\psi}\frac{\partial^2}{\partial x_i^2}\psi$
& Kinetic energy\\
$J = \int G(\rho )$ & Self interaction energy
\end{tabular}
\label{moment-definitions}
\end{table}

\end{document}